\documentclass[prl,twocolumn,preprintnumbers,amsmath,amssymb,superscriptaddress]{revtex4-2}
\usepackage{graphicx}
\usepackage{subcaption} 
\usepackage{xcolor}
\usepackage{svg}
\usepackage{dcolumn}
\usepackage{bm}
\usepackage{hyperref}
\usepackage{amsmath}
\begin{document}
\title{First-passage statistics of confined colloids}
\author{Guirec de Tournemire}
\author{Nicolas Fares}
\author{Yacine Amarouchene}
\email{yacine.amarouchene@u-bordeaux.fr}
\author{Thomas Salez}
\email{thomas.salez@cnrs.fr}
\affiliation{Univ. Bordeaux, CNRS, LOMA, UMR 5798, F-33405 Talence, France.}
\date{\today}
\begin{abstract}
The encounter of diffusing entities underlies a wide range of natural phenomena. The dynamics of these first-passage dynamics are strongly influenced by confining geometries. Confinement modifies microscopic diffusion through conservative and hydrodynamic interactions, making it essential for realistic modeling. In this Letter, we investigate how confinement affects the first-passage statistics of a diffusing particle.
Using \textit{state-of-the-art} holographic microscopy combined with advanced statistical inference, we probe this motion with nanometric precision. Our experimental and numerical results show that confinement can either hinder or enhance first-passage kinetics, depending on the spatial direction. In particular, wall-normal target finding is accelerated by confinement-induced non-Gaussian displacement statistics, which increases the probability of rare, large displacements, with implications for confined chemical reactions and biological \textit{winners-take-all} processes near boundaries.
\end{abstract}
\maketitle

\textbf{Significance statement:} 
\textit{Stochastic first-passage problems are ubiquitous in physics, chemistry, and biology. Indeed, winners-take-all processes such as proteins finding the codons on DNA, neurotransmitters crossing the synaptic gap, or chemical reactants forming intermediate high-energy transition states, rely on first-passage dynamics, and are governed by rare events. In addition, confining boundaries and surface effects are inherent to such realistic situations. Remarkably, near boundaries, three main effects -- hydrodynamic interactions, conservative surface interactions, and resulting non-Gaussian displacements -- arise and modify drastically Brownian motion. A natural but open question emerges on whether and how these confinement effects impact the target-finding dynamics. In this study, we combine state-of-the-art experiments in order to track colloids near walls with high resolution, and we quantitatively address this question.}

The efficiency of stochastic target finding, where noise-driven particles locate specific sites, is fundamental to diverse natural and engineered systems. Examples range from chemical and prebiotic reactions~\cite{gillespie1977exact,allen1980brownian}, the transfer of information in neuronal synapses~\cite{heine2008surface,armbruster2020effects,reva2021first}, DNA-protein interactions~\cite{shin2019target,lucas20143d,zhang2016first, valverde2025macromolecular}, or even Stock-market predictions~\cite{chicheportiche2014some,patie2004some}. In such processes, micro- or nano-sized entities moving in a viscous liquid typically undergo Brownian motion, a diffusion process driven by thermal fluctuations ~\cite{einstein1905brownian,smoluchowski1906kinetic,perrin1910mouvement,li2010measurement}. The dynamics and success of these searches are profoundly influenced by the complex confining geometries of the local environments, such as blood cells flowing in vessels ~\cite{seiffge1986passage,doster2007microscopic}, transmitter and receptor proteins in synaptic junctions~\cite{heine2008surface,armbruster2020effects}, or ions in chemical solutions~\cite{allen1980brownian}. Such biological or chemical environments are inherently complex and significantly modulate particle diffusivity and overall behavior.

In recent decades, environment-dependent particle dynamics has fostered extensive research efforts, in both idealized and more realistic scenarios. For example, it has been shown that non-Newtonian polymeric liquids exhibit non-Markovian dynamics~\cite{jeney2008anisotropic,guerin2016mean}. In the case of purely viscous fluids, confining boundaries are known to hydrodynamically hinder ~\cite{faucheux1994confined} or speed up~\cite{joly2006probing,alexandre2022stickiness} diffusion, depending on the boundaries' wetting and stickiness properties ~\cite{burada2009diffusion}. Such systems exhibit anisotropic spatial variations in diffusivity~\cite{eral2010anisotropic}. 
In addition, diffusing entities can be subjected to electrostatic or Van der Waals forces~\cite{bouzigues2008nanofluidics} originating from confining boundaries, or other diffusing entities~\cite{sainis2007statistics,valverde2025macromolecular}.
Interestingly, the combination of confinement-induced static and hydrodynamic interactions often lead to non-Gaussian dynamics~\cite{guan2014even,matse2017test,alexandre2023non} - in  stark contrast with the Gaussian dynamics of unconfined, bulk systems. Consequently, and given the importance and ubiquity in biology and chemistry of target-finding situations near boundaries, it is essential to study how the properties of a confining environment influence the problem of target finding. This problem is fundamentally probabilistic and is analyzed using the statistics of the first-passage time (FPT)-the random time it takes a diffusively searching entity to first reach a specific target~\cite{redner2001guide,grebenkov2020preface}. Despite extensive studies on FPT statistics in diverse systems~\cite{reva2021first,grebenkov2020preface,guerin2012non,dolgushev2025evidence,mattos2012first,sposini2018first,besga2020optimal}, the fundamental case involving classical static and hydrodynamic interactions with a single wall remains unexplored. This work directly addresses this gap.

Experimentally monitoring rare events like colloidal target finding is challenging, as it demands a technique that spans broad spatial and temporal scales. To overcome this obstacle, we employ here holographic microscopy to capture three-dimensional trajectories of isolated colloidal particles diffusing in a viscous liquid near a planar wall. Statistical inference of these trajectories reveals non-Gaussian displacement distributions, indicating a heightened probability of large, rare displacements. These confinement-induced non-Gaussian dynamics appear to accelerate FPT processes, hence offering a new perspective on target-reaching scenarios in realistic, bounded environments.

\begin{figure}[t!]
    \centering
    \includegraphics[width=0.98\linewidth]{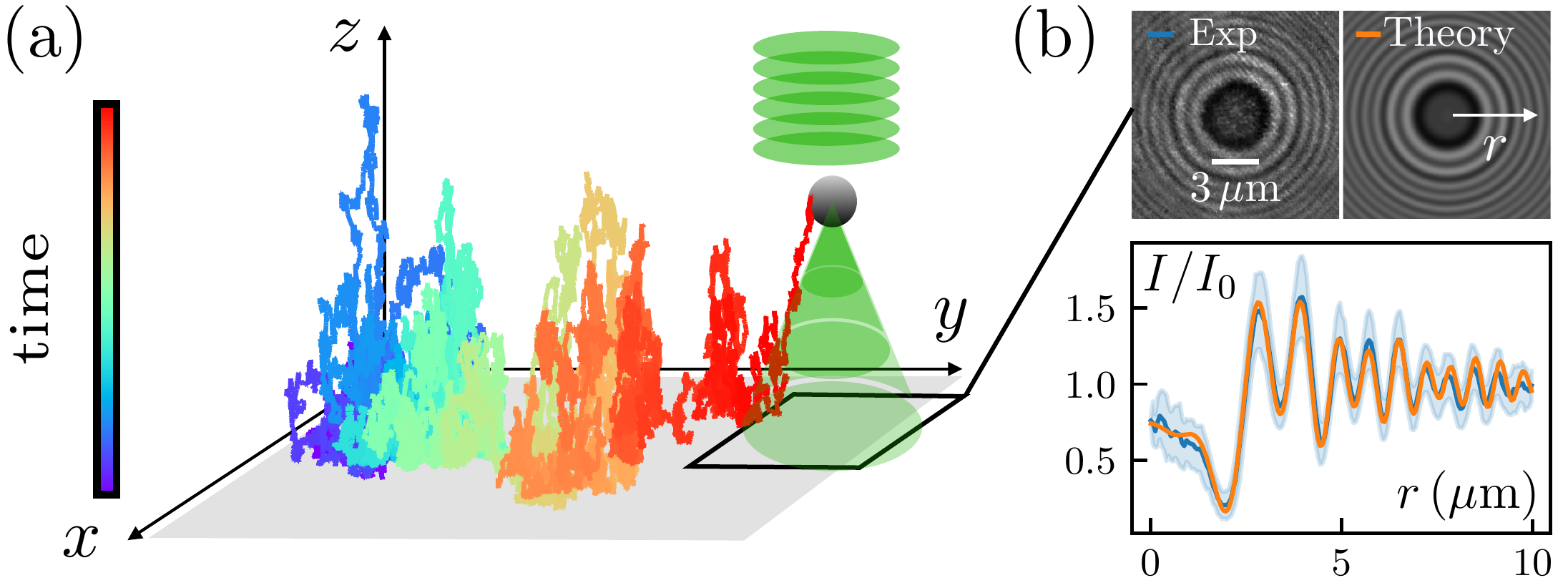}\\
    \includegraphics[width=1\linewidth]{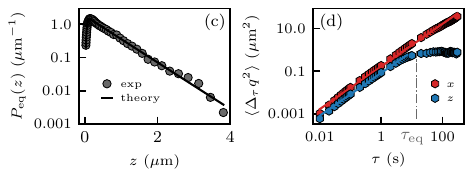}\\
    \includegraphics[width=1\linewidth]{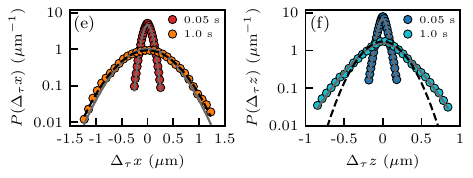}
    \caption{\small Monitoring confined diffusion via Mie holography~\cite{lavaud2021stochastic}. 
    (a) Three-dimensional trajectory of a spherical polystyrene colloid of radius $a_\mathrm{p}=1.48 \,\mathrm{\mu m}$ diffusing in a viscous liquid near a rigid glass wall.  (b) Raw experimental (Exp) hologram, and corresponding theoretical hologram, with their respective radial intensity profiles. 
    (c) Experimental (exp) equilibrium probability distribution function, $P_\mathrm{eq}$, for the altitude $z$ of the colloid (gray circles), with its corresponding fit to Eq.~\ref{eq:Peq} (black line), leading to $B = 6.96$, $l_\mathrm{D} = 33 \, \mathrm{nm}$, and $l_\mathrm{B} = 604 \, \mathrm{nm}$. (d) Mean Square Displacements $\langle \Delta_\tau q^2\rangle$ as functions of the time lag $\tau$, in the $x$-direction (red) and the $z$-direction (blue). The corresponding colored solid lines depict the best fits to Eq.~\ref{eq:MSD} in both directions, leading to $D_x = 0.542\, D_0$ and $D_z = 0.268\, D_0$. The equilibrium time is indicated with the dot-dashed vertical line. 
    (e)~Probability Distribution Function, $P(\Delta_\tau x)$, of the colloid's in-plane displacements $\Delta_\tau x$ over a time $\tau$, for $\tau = 50\,$ms (red dots) and $1\,$s  (orange dots). The gray solid lines depicts the best fits to the theoretical prediction detailed in Refs.~\cite{lavaud2021stochastic,alexandre2023non}, with the same parameter values as in panel~(c). The black dashed line corresponds to the Gaussian prediction of Eq.~\ref{eq:gauss}, with the diffusion coefficient $D_x$ from panel~(d). 
    (f) Same as panel~(e) but in the $z$-direction.}
    \label{fig:figure_1}
\end{figure}

\vspace{0.3cm}
\textit{Methods---}
In this Letter, the confined diffusion we refer to consists of the Brownian motion of an isolated negatively surface-charged spherical polystyrene colloid, of radius $a_\mathrm{p}$ (ranging from $1\,\mathrm{\mu m}$ to $3\,\mathrm{\mu m}$) and density $\rho_\mathrm{p} \simeq 1050 \, \mathrm{kg/m^3}$, diffusing in a viscous water-ethylene-glycol mixture of density $\rho_\mathrm{m}$ and viscosity $\eta_\mathrm{m}$ (ranging from $1000\,\mathrm{kg/m^3}$ to $1050\,\mathrm{kg/m^3}$ and from $0.95\,\mathrm{mPa\cdot s}$ to $3.2\,\mathrm{mPa\cdot s}$, respectively), and near a negatively surface-charged rigid glass wall, as illustrated in~Fig.~\ref{fig:figure_1}a. 

The microscopy method used to monitor the confined diffusion is based on Lorenz-Mie holography~\cite{lee2007characterizing,lavaud2021stochastic,fares2024observation}, and is schematized in Fig.~\ref{fig:figure_1}a. The sample preparation protocol is detailed in Supplementary Information (\textbf{SI} Sec. 8). A plane wave (wavelength 532~nm; output power 4~mW) illuminates the colloid, which scatters part of the incident wave. The interference between the incident and scattered waves produces fringes, that recast in a pattern called a hologram. An experimental hologram is displayed in~Fig.~\ref{fig:figure_1}b.
The experimental hologram is then fitted to the Lorenz-Mie theory, which leads to the determination of the colloid's radius and refractive index $n_\mathrm{p}$ (allowing for \textit{in-situ} calibration of each experiment), and of the colloid's three-dimensional position with nanometric precision, as suggested by the agreement between the experimental and theoretical holograms and associated radial intensity profiles represented in~Fig.~\ref{fig:figure_1}b. 

In practice, holograms are recorded \textit{via} a 100x Olympus immersion-oil objective and a Basler camera, operating at a frame rate  of 100 frames per second with an exposure time of 2 ms, over a total recording duration of one hour. The reconstructed colloid's trajectory $(x(t),y(t),z(t))$ (see~Fig.~\ref{fig:figure_1}a) is then analyzed in order to infer characteristic features of confined diffusion, with $(x,y)$, $z$ and $t$ denoting respectively the colloid's in-plane position, the colloid-wall gap distance and the time. 

The first feature described here stems from the $z$-trajectory and allows for a quantitative description of the confinement considered. By binning the $z$-domain and counting positional occurencies at long times, one gets the equilibrium probability distribution function $P_\mathrm{eq}$ of presence in the $z$-direction, as displayed in~Fig.~\ref{fig:figure_1}c. 
While the electrostatic repulsion between the negatively-charged colloid and wall sets the lower boundary of the $z$-domain, an effective upper boundary is set by the colloid's buoyant weight.
Specifically, $P_\mathrm{eq}$ directly depends on to the potential $U_\mathrm{eq}$ governing the process, as:
\begin{equation}
    P_\mathrm{eq}(z) = \mathcal{N} \mathrm{e}^{-U_\mathrm{eq}(z) / (k_\mathrm{B} \Theta)} \, ,
    \label{eq:Peq}
\end{equation} where $\mathcal{N}$ is a normalization constant, and where $k_\mathrm{B}$ and $\Theta$ denote the Boltzmann constant and room temperature, respectively. The potential $U_\mathrm{eq}$ reads:
\begin{equation}
    \frac{U_\mathrm{eq}}{k_\mathrm{B} \Theta} = B \mathrm{e}^{-z/l_\mathrm{D}} + \frac{z}{l_\mathrm{B}}.
    \label{eq:Ueq}
\end{equation}
In the above equation, $B$ quantifies the magnitude of surface charges~\cite{behrens2001charge}, $l_\mathrm{D}$ is the Debye length~\cite{sze2003zeta}, \textit{i.e.} the ionic screening length of the electrostatic repulsion, and $l_\mathrm{B} = \frac{k_\mathrm{B} \Theta}{\frac{4}{3}\pi a_\mathrm{p}^3\Delta \rho g}$ (with $\Delta \rho = \rho_\mathrm{p} - \rho_\mathrm{m}$ the density mismatch and $g$ the local gravity constant) is the Boltzmann length~\cite{perrin1910mouvement}, \textit{i.e.} the typical extent of the reachable $z$-positions.
In fact, we scan the confinement strength \textit{via} the Boltzmann length, by adding ethylene glycol to water solutions and thus tuning the density mismatch (see \textbf{SI} Sec.2).

We now focus on the confined-diffusion dynamics. The canonical observable related to diffusion, namely the Mean Squared Displacement (MSD) $\langle \Delta_\tau q^2 \rangle$ (where  $q = x, y,z$, $\Delta_\tau q = q(t+\tau) - q(t)$, and $\langle\cdot\rangle$ denotes the average over the time $t$), is shown in~Fig.~\ref{fig:figure_1}d. The process is anisotropic and the diffusivity is lower in the normal direction than in the transverse one, as $z$-data points are systematically lower than $x$-data points. Nevertheless, at short times, both MSDs are still linear with respect to time, as in bulk configurations, and are ruled by the classical diffusion law: 
\begin{equation}
    \label{eq:MSD}
    \langle \Delta_\tau q ^2 \rangle  = 2 D_q \:\tau,
\end{equation}where $D_q$ is the diffusion coefficient of the process. 
In the absence of a wall, this diffusion coefficient would be the bulk one $D_0$, that obeys the Stokes-Einstein relation:
\begin{equation}
    D_0 = \frac{k_\mathrm{B}\Theta}{\gamma}, 
    \label{eq:D0}
\end{equation} where $\gamma = 6 \pi \eta_\mathrm{m} a_\mathrm{p}$ is the bulk Stokes drag coefficient. 
However, the confining wall, through the associated no-slip boundary condition, amplifies the viscous drag and makes it not only anisotropic but also $z$-dependent~\cite{faxen1923bewegung,brenner1961slow}. Consequently, the colloid's real diffusivity $d_q(z)$ becomes $z$-dependent (see \textbf{SI} Sec.1), leading to the $z$-averaged diffusion coefficient $D_q$, given by ~\cite{lavaud2021stochastic}:
\begin{equation}
    D_q =  \int_0^{+ \infty} \text{d}z\, P_\mathrm{eq}(z) d_q(z) .
    \label{eq:D_avg}
\end{equation} 
The average diffusion coefficient $D_q$ is thus reduced compared to $D_0$.
Besides, we stress here that the $z$-MSD saturates at longer times, to a value that scale as $l_\mathrm{B}^2$~\cite{lavaud2021stochastic}. This effective trapping directly results from the potential $U_\mathrm{eq}$ in which the colloid lies. This saturation allows to define the characteristic equilibration time $\tau_\mathrm{eq} \sim l_\mathrm{B}^2 / (2D_z)$ of the process, which is the time scale over which the colloid has visited the typically-reachable $z$-positions. 

Our statistical inference goes further than the simple MSDs, \textit{i.e.} the variances of the displacement distributions, as one can in fact reconstruct the full distributions $P(\Delta_\tau x)$ and $P(\Delta_\tau z)$ for, respectively, in-plane and normal displacements over a duration $\tau$. For both these observables, which are displayed in~Figs.~\ref{fig:figure_1}e,f, we observe a broadening over time, highlighting the diffusive nature of confined Brownian motion. Moreover, for our parameter range, and despite the presence of non-Gaussianities~\cite{lavaud2021stochastic,alexandre2023non} due to a multiplicative noise, the in-plane distribution can be fairly approximated by a Gaussian law, $P_\mathrm{Gauss}$, that reads:
\begin{equation}
    \label{eq:gauss}
    P_\mathrm{Gauss}(\Delta_\tau q) = \frac{1}{\sqrt{4\pi D_q \, \tau}} \: \mathrm{e}^{ \frac{-\Delta_\tau q^2}{4D_q\,\tau}}.
\end{equation}
As such, the in-plane process resembles a simple slowed-down ($D_x \leq D_0$) bulk-like Gaussian process. In sharp contrast, we observe the emergence of strong non-Gaussian tails in the $z$-displacement distribution, mostly due to the trapping potential -- another clear signature of the confinement-induced symmetry breaking between spatial directions.
This feature has been extensively described, and recasts into a \textit{Brownian-yet-non-Gaussian} process~\cite{guan2014even,alexandre2023non,millan2023numerical}. However, a central question emerges for our purpose here: Do such non-Gaussianities, associated with an increased probability of the rarest events, impact the FPT statistics ? This is what we quantitatively address in the following. 

\vspace{0.4cm}
\textit{Results---}
Having assessed the reference bulk case in the \textbf{SI} Sec.5, we turn to a detailed examination of how confinement influences the first-passage statistics, through the first-passage time distribution (FPTD) $f(t)$ defined as the distribution of times at which the particle first reaches a target located at a distance $L$ from its initial position.

As mentioned previously, the confinement-induced hydrodynamic interactions alter the colloid’s mobility, both perpendicular~\cite{brenner1961slow} and parallel~\cite{faxen1923bewegung} to the wall. Hence, the stochastic processes are treated starting from the following set of coupled Langevin equations: 
\begin{equation}
\begin{cases}
\begin{aligned}
m\ddot{x}(t) &= -\gamma_x(z)\dot{x}(t) + \sqrt{2 k_\mathrm{B} \Theta \gamma_x(z)} \,\xi_x(t), & (\text{a}) \\
m\ddot{z}(t) &= -\gamma_z(z)\dot{z}(t) - U_\mathrm{eq}'(z) + \sqrt{2 k_\mathrm{B} \Theta \gamma_z(z)} \,\xi_z(t), & (\text{b})
\end{aligned}
\end{cases}
\label{eq:Langevin_confined}
\end{equation}
where $\gamma_x(z)$ and $\gamma_z(z)$ are the $z$-dependent viscous drag coefficients along $x$~\cite{faxen1923bewegung} and $z$~\cite{brenner1961slow}, respectively, where $\xi_x(t)$ and $\xi_z(t)$ are independent Gaussian white noises, satisfying $\langle\xi_q\rangle = 0$ and $\langle \xi_q(t)\xi_q(t')\rangle = \delta(t-t')$, and where $U_\mathrm{eq}'$ denotes the $z$-derivative of the potential $U_\mathrm{eq}(z)$ (see Eq.~\ref{eq:Ueq}).

In contrast with the aforementioned bulk scenario, given the wall-induced anisotropy in the problem, we distinguish between two target-finding processes, namely a parallel search along the $x$-axis (the same process can be discussed along $y$), or a normal search along the $z$-axis. 
The aim of the two following subsections is to characterize the FPTD of the confined colloid in these two situations.  

\begin{figure}[t!]
    \centering
    \includegraphics[width=0.9\linewidth]{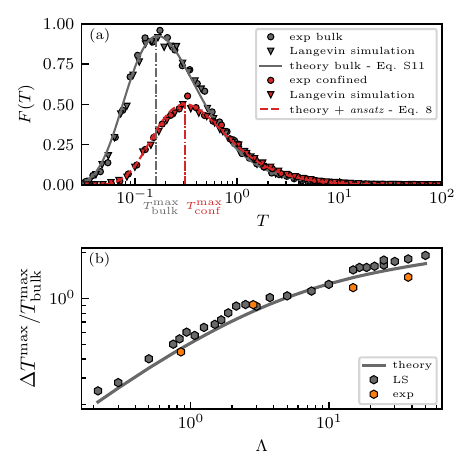}
    \caption{\small First-passage-time statistics in confinement for a wall-parallel search with a target located at a distance $L = 500 \, \mathrm{nm}$ from the initial position. (a) Dimensionless distributions $F(T)$ of the dimensionless first-passage times $T$ of a colloid diffusing, either in the bulk (gray) or near a wall (red). Circles correspond to experimental (exp) data, upside-down triangles to Langevin simulation (LS) data (see Eq.~\ref{eq:Langevin_confined}a), and the solid lines to the theoretical predictions of Eq. (S11) (bulk, $D_0 = 4.53.10^{-2} \, \mathrm{\mu m^2 / s}$) and Eq.~\ref{eq:ansatz} (confined, $D_0 = 15.1.10^{-2}\, \mathrm{\mu m^2 / s}$, $D_x = 8.18.10^{-2} \, \mathrm{\mu m^2 / s}$). (b) Relative delay $\Delta T^\mathrm{max} / T_\mathrm{bulk}^\mathrm{max}$ (see Eq.~\ref{rtd}) of the most-likely dimensionless first-passage time as a function of the confinement parameter $\Lambda = a_\mathrm{p} / l_\mathrm{B}$. Orange hexagons correspond to experiments, gray hexagons to simulations, and the gray line to the associated theoretical prediction (see \textbf{SI} Sec.6), with no adjustable parameter.}
    \label{fig:figure_2}
\end{figure}

In the in-plane dynamics described by Eq.~\ref{eq:Langevin_confined}a, hydrodynamic interactions with the wall constitute the only deviation from the bulk behavior (Eq. S7 of the \textbf{SI}). The in-plane motion exhibits a MSD that increases linearly with time (Fig.~\ref{fig:figure_1}d), as well as a nearly-Gaussian displacement distribution (Fig.~\ref{fig:figure_1}e) for our current range of parameters~\cite{lavaud2021stochastic,alexandre2023non}), both fully characterized by the average diffusion coefficient $D_x$. 
Accordingly, the confined in-plane dynamics -- characterized by a $z$-dependent diffusivity $d_x(z) = k_\mathrm{B}\Theta / \gamma_x(z)$ -- can be fairly approximated by a bulk-like Gaussian process but with a lower, effective diffusivity $D_x$. As a consequence, we substitute $D_x$ into the Lévy distribution of Eq. S11 originally derived for the bulk case, which then takes the following form:
\begin{equation}
    f(t) = \frac{L}{\sqrt{4 \pi D_x t^3}} \text{e}^{-\frac{L^2 }{4 D_x t}},
    \label{eq:ansatz}
\end{equation} 
and stands as a theoretical \textit{ansatz} for the in-plane target-finding process.
As shown in Fig.~\ref{fig:figure_2}a, Eq.~\ref{eq:ansatz} correctly captures the experimental FPTDs. 
This agreement is further confirmed by numerical Langevin simulations mimicking the experimental confined diffusion, following the method of Ref.~\cite{millan2023numerical} (see \textbf{SI} Sec.4 for details on the simulations). 
Variables are rescaled here in order to account for the different viscous liquids considered, through: $T = D_0t/L^2$ and $F(T) = f(t) \frac{L^2}{D_0}$. 

Since the mean FPT is infinite in the unbounded wall-parallel domain, we focus on the most-likely FPT, denoted as $T^\mathrm{max}$. The latter increases due to confinement, as qualitatively highlighted in Fig.~\ref{fig:figure_2}a. Intuitively and as discussed above, the colloid is essentially slowed down by wall-induced friction as it moves near and parallel to the wall, thus postponing the typical moment in time at which it can reach a given target. In order to quantitatively evaluate the magnitude of such an effect, we define a relative time delay with respect to the bulk reference, as: 
\begin{equation}
    \frac{\Delta T^\mathrm{max} }{ T_\mathrm{bulk}^\mathrm{max} } = \frac{T_\mathrm{conf}^\mathrm{max} - T_\mathrm{bulk}^\mathrm{max}}{T_\mathrm{bulk}^\mathrm{max}},
    \label{rtd}
\end{equation} 
in which $T_\mathrm{bulk}^\mathrm{max}$ and $T_\mathrm{conf}^\mathrm{max}$ denote the most-likely FPT in bulk and confined situations, respectively. 
As shown in Fig.~\ref{fig:figure_2}b, the relative time delay -- deduced from FPTDs of both the experiments and Langevin simulations -- increases with the confinement parameter $\Lambda  = \frac{ a_\mathrm{p} }{ l_\mathrm{B}}$. The experimental and simulated data are consistent with the theoretical prediction (see details in \textbf{SI} Sec.6). We stress that, in normalized units, the value of the distance $L$ to the target location does not affect this result. In summary, the confined wall-parallel target-finding process is bulk-like, but with wall-induced friction slowing down the clock.

\begin{figure}[t!]
    \centering
    \includegraphics[width=0.9\linewidth]{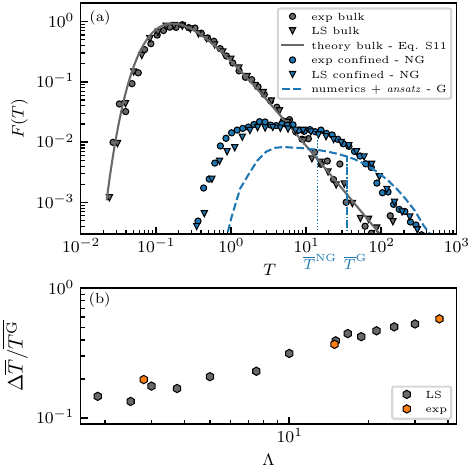}
    \caption{\small First-passage-time statistics in confinement for a target located along $z$ at a distance $L = 300 \, \mathrm{nm}$ from the initial position. Note that the latter is randomly selected from $P_{\textrm{eq}}(z)$ for pre-thermalized samples. (a) Dimensionless distribution $F(T)$ of the dimensionless first-passage times $T$ for a colloid diffusing either in the bulk (gray) with $D_0 = 4.53.10^{-2} \, \mathrm{\mu m^2 / s}$, or near a wall (blue) with $D_0 = 7.14.10^{-2}\mathrm{\mu m^2 / s}$ and $D_z = 4.3.10^{-3} \mathrm{\mu m^2 / s}$. Circles correspond to experimental data (non-Gaussian), upside-down triangles to simulation data (non-Gaussian) (see Eq.~\ref{eq:Langevin_confined}b), the gray solid line to the bulk theoretical prediction of Eq. S11, and the blue dashed line to simulation data when replacing the $z$-dependent diffusivity $d_z(z)=k_{\textrm{B}}\Theta/\gamma_z(z)$ by $D_z$ within Eq.~\ref{eq:Langevin_confined}b (Gaussian).
    (b) Relative reduction $\Delta \overline{T} / \overline{T^\mathrm{G}}$ (see Eq.~\ref{eq:MFPT}) of the dimensionless mean first-passage time as a function of the confinement parameter $\Lambda = a_\mathrm{p} / l_\mathrm{B}$. Langevin simulations (LS) are indicated with gray hexagons, while experiments (exp) are indicated with orange hexagons.}
    \label{fig:figure_3}
\end{figure}

\vspace{0.5cm}
We now turn to the wall-normal search dynamics, along the $z$-axis. 
The $z$-process exhibits a markedly richer dynamics (see Eq.~\ref{eq:Langevin_confined}b), that strongly affects the FPTD, as discussed hereafter. 
First, as evidenced in Fig.~\ref{fig:figure_2}a \textit{via} the overall time shift between the bulk and confined FPTDs, the FPT dynamics is essentially slowed down as compared to the corresponding bulk FPT dynamics, which echoes the ``slowed-clock'' effect observed along $x$. 
However, in the confined scenario, the $z$-FPTD exhibits a noticeable flattening, which stems from the combined influence of gravity and electrostatic repulsion, through the equilibrium potential $U_\mathrm{eq}$ (see Eq.~\ref{eq:Ueq}). 
Such a static origin of the FPTD flattening can be demonstrated by coarsely replacing the $z$-dependent diffusivity $d_z(z)=k_{\textrm{B}}\Theta/\gamma_z(z)$ by the constant, average one $D_z$ within Eq.~\ref{eq:Langevin_confined}b, and then numerically solving the latter. Then, by constructing the FPTD from the simulated trajectories, one indeed sees a functional form that is flattened (see dashed line in Fig.~\ref{fig:figure_3}a), as compared to the bulk FPTD.  

Furthermore, and more strikingly, the aforementioned coarse numerical simulations (blue dashed line, with $D_z$ replacing $d_z(z)$ making the process Gaussian : ``-G'') , while properly capturing the flattening of the FPTD, fail to quantitatively capture the experimental and full numerical simulations (blue circles and triangles, respectively, with $d_z(z)$ triggering non-Gaussianities : ``-NG''). Such a discrepancy is rooted in the confinement-induced hydrodynamic interactions with the wall, which are more pronounced along $z$ than along $x$. Indeed, while $\gamma_x(z)$ scales as $\log{(a/z)}$~\cite{chaoui2003creeping} and is nearly a plateau in the accessed $z$-range~\cite{faxen1923bewegung,lavaud2021stochastic}, $\gamma_z(z)$ scales as $a/z$~\cite{brenner1961slow,lavaud2021stochastic}. The latter space-dependent drag coefficient along $z$, coupled to the potential $U_{\textrm{eq}}(z)$ leads to large non-Gaussianities of the $z$-displacement probability (see Fig.~\ref{fig:figure_1}f). Consequently, the probability of large displacements is increased, as compared to a Gaussian reference case (such as the one in the coarse simulations), and the search process gets comparatively faster.

Since the mean FPT (MFPT) $\overline{T}$ ($\overline{\cdot}$ denoting the average weighted by $F(T)$) is finite in the wall-normal $z$-domain, due to the confining potential $U_{\textrm{eq}}(z)$, one can use it as a relevant measure of the typical search time in this direction. We compare the experimental or numerical MFPT $\overline{T^\mathrm{NG}}$, which includes non-Gaussian effects, with the MFPT $\overline{ T^\mathrm{G}}$ obtained from the coarse simulations described above in which such non-Gaussian effects are suppressed by construction.
To do so, we introduce a relative time reduction, expressed as:
\begin{equation}
    \frac{\Delta \overline{T}}{\overline{T^\mathrm{G}}} =\frac{\overline{T^\mathrm{G}}  - \overline{T^\mathrm{NG}}}{ \overline{T^\mathrm{G}}}.
    \label{eq:MFPT}
\end{equation} As shown in Fig.~\ref{fig:figure_3}b, this relative time reduction is positive and increases with confinement. We stress that, in normalized units, the value of the distance $L$ to the target does not qualitatively affect this result (see \textbf{SI} Sec.7). In summary, confinement-induced non-Gaussianities boost the target-finding process in the wall-normal direction.

\vspace{0.4cm}
\textit{Discussions---}
To conclude, we have investigated first-passage processes in presence of confinement effects, such as surface forces, hydrodynamic interactions with a wall, and resulting non-Gaussian displacements. By the use of holographic microscopy and statistical inference, usual observables related to Brownian motion, such as displacement distributions, mean-square displacements and equilibrium distributions in position, were measured with state-of-the-art precisions allowing for a quantitative analysis of first-passage stochastic dynamics in confinement. Our main findings, backed up by theoretical and numerical models, are listed hereafter. First, if the colloid randomly moves transversely to the wall, the process remains almost Gaussian and bulk-like, but the clock is slowed down as compared to the bulk case due to the wall-induced friction. Hence, target finding is slowed down as well. Second, if the colloid moves in the direction normal to the wall, the process is strongly non-Gaussian due to the intricate combination of wall-normal forces and space-dependent mobility. Strikingly, target finding is then typically re-accelerated by such non-Gaussianities, compared to a Gaussian process under the same potential with a hindered mobility. This key result can be understood by recalling that the probability of observing large displacements is enhanced in this case through the presence of the non-Gaussian tails in the displacement distribution. Such a non-Gaussianity-induced enhanced target-finding dynamics of confined colloids opens promising paths for confined chemistry~\cite{grommet2020chemical}, and biophysics~\cite{fakhri2010brownian}, and might have drastic consequences in winners-take-all processes near boundaries, such as fecundation, where one single rare event can entirely dominate the dynamics~\cite{sposini2024being}. Investigating further the influence of soft boundaries in such problems, through Brownian elastohydrodynamic couplings~\cite{Ye2025, Lacherez2025,mcgraw2025transport,Mohammadi2025}, would be a relevant task for the future. 

\vspace{0.3cm}
\textit{Data availability---}
Experimental data and analysis codes have been deposited in \url{https://github.com/EMetBrown-Lab/first-passage-time/}.

\vspace{0.3cm}
\textit{Acknowledgements---}
The authors thank Elodie Millan and Maxime Lavaud for useful insights into materials, as well as Juliette Lacherez, Leah Anderson, Joshua McGraw and David Dean for enlightening discussions. They also thank Josiane Parzych and Val\'erie Thouard for financial project management, as well as Th\'eo Guillaume, Romain Houques and Anne Tempel for lab-space installation. Moreover, they acknowledge financial support from the European Union through the European Research Council under EMetBrown (ERC-CoG-101039103) grant. Views and opinions expressed are however those of the authors only and do not necessarily reflect those of the European Union or the European Research Council. Neither the European Union nor the granting authority can be held responsible for them. The authors also acknowledge financial support from the Agence Nationale de la Recherche under EMetBrown (ANR-21-ERCC-0010-01), Softer (ANR21-CE06-0029), and Fricolas (ANR-21-CE06-0039) grants, as well as from the Interdisciplinary and Exploratory Research program at Univ. Bordeaux under MISTIC grant, France. Besides, they acknowledge the support from the LIGHT Sciences and Technologies Graduate Program (PIA3 Investment for the Future Program, ANR-17EURE-0027), and the Réseau de Recherche Impulsion (RRI) “Frontiers of Life”, which received financial support from the French government in the framework of Univ. Bordeaux’s France 2030 program. Finally, they thank the Soft Matter Collaborative Research Unit, Frontier Research Center for Advanced Material and Life Science, Faculty of Advanced Life Science at Hokkaido University, Sapporo, Japan, and the CNRS International Research Network between France and India on ``Hydrodynamics at small scales: from soft matter to bioengineering''.
\bibliography{Tournemire2025.bib}
\end{document}